\documentclass[aps,prl,twocolumn,showpacs,preprintnumbers,amsmath,amssymb,amsfonts]{revtex4}
\usepackage{graphicx}
\usepackage{textcomp}
\usepackage{color}

\usepackage{times}

\begin{document}

\title{Complete temporal characterisation of asymmetric pulse compression in a laser wakefield}

\author{J.~Schreiber$^{1}$}
\email[]{j.schreiber@imperial.ac.uk}
\author{C.~Bellei$^{1\dag}$}
\author{S.~P.~D.~Mangles$^{1}$}
\author{C.~Kamperidis$^{1\ddag}$}
\author{S.~Kneip$^{1}$}
\author{S.~R.~Nagel$^{1}$}
\author{C.~A.~J.~Palmer$^{1}$}
\author{P.~P.~Rajeev$^{2}$}
\author{Z.~Najmudin$^{1}$}

\affiliation{$^{1}$ Blackett Laboratory, Imperial College, London SW7
2AZ, United Kingdom}
\affiliation{$^{2}$ Central Laser Facility, Rutherford-Appleton
Laboratory, Chilton, Oxon, OX11 0QX, United Kingdom}

\date{\today}

\begin{abstract}
We present complete experimental characterisation of the temporal shape of an intense ultrashort
200-TW laser pulse driving a laser wakefield. The phase of the pulse was uniquely
measured using (second order) frequency
resolved optical gating (FROG). The pulses are
asymmetrically compressed, and exhibit a positive chirp consistent with the expected asymmetric self-phase modulation due to photon acceleration/deceleration in a relativistic plasma wave. The measured pulse duration decreases linearly with increasing length and density of the plasma, in quantitative agreement with the intensity dependent group velocity variation in the plasma wave.
\end{abstract}


\maketitle

Laser wakefield accelerators have now demonstrated the production of quasimonoenergetic electron bunches to GeV-scale energies over only $\sim$ cm lengths \cite{LeemansNatPhys, KneipPRL2009}. 
Though remarkably short for an accelerator, these distances are still many times the Rayleigh length, $z_{R}$, of  the driving high-intensity laser pulse, so that the non-linear interplay between laser and plasma becomes important. This leads to a wealth of interesting phenomena such as self-focusing \cite{selffoc, thomas}, self-phase modulation \cite{Mori97, watts} or photon acceleration \cite{photon} and pulse shortening \cite{gordon, FaurePRL2005}. In combination, these processes can result in $a^{2}$ (intensity) amplification of the laser pulse \cite{gordon, KneipPRL2009}. 
Here $a = eE/(m_e\omega_{0} c)$ is the normalised vector potential for a laser with electric field $E$ and angular frequency $\omega_{0}$. 
Laser wakefields are driven by the ponderomotive force of the laser, $F_{p} = -\frac{1}{\gamma}m_ec^{2}\nabla a^{2}$, where $\gamma^{2} =1+{a^2}/{2}$ is the relativistic factor due to the transverse quiver of electrons in the laser field. Hence, determining the evolution of $a^{2}$ is of vital importance in understanding wakefield accelerators.
For example, $a^{2}$ amplification was essential in the first demonstrations of monoenergetic beam production \cite{mono}.  

For the interaction to extend to many Rayleigh lengths, the laser must be prevented from diverging naturally. This can be through the action of an external guiding structure \cite{LeemansNatPhys}, or simply through a combination of ponderomotive and relativistic self-focusing \cite{KneipPRL2009}. For sufficiently intense ultrashort laser pulses ($a>1$), the laser will quickly focus to a matched spot radius, $w_{m} \approx \frac{1}{\pi}\sqrt{a_{0}}\, \lambda_{p}$ where $\lambda_{p} \equiv 2\pi c/\omega_{p}$, 
$\omega_p$ is the plasma frequency
and $a_{0}$ is the peak vacuum value of $a$ \cite{lu, thomas}. Once matched, transverse variations in spot size play little role in increasing $a^{2}$. 

However, as the laser sits in the density variation created by the plasma wave, it experiences a spatiotemporally varying refractive index 
$\eta = \left( 1 - \omega_{p}^{2}/(\gamma \omega_{0}^{2}) \right)^{1/2}$. 
For the front of the plasma wave, the density increase as plasma electrons are pushed forward is mostly compensated by the intensity dependence of $\eta$ (through the dependence on $\gamma$) \cite{sprangle}.
However, beyond this, both density decrease and further increase of $\gamma$ cause $\eta$ to increase. This results in an increase in group velocity $v_{g} \approx \eta c$, and thus compression of the pulse increasing towards the back of the plasma wave. 

Associated with pulse compression must be an increase in spectral bandwidth. At the front of the pulse, density and intensity variations cancel to limit the amount of frequency shifts. However beyond the first maximum of the plasma wave, both decreasing density and increasing intensity cause an increasing $\eta$ and thus a reciprocal decrease in phase velocity $v_{\varphi}$. Hence, the majority of the pulse can be redshifted. For longer pulses ($c\tau \approx \lambda_{p}$), both increasing plasma density and relativistic self-phase modulation (decreasing $\gamma$) act together to cause strong blueshifts at the back of the pulse. 

At high intensity and including multidimensional effects, this picture is complicated further. The plasma wave becomes non-linearly steepened, and the centre of the plasma wave evolves into a caviton-like structure with relatively small variations in $\eta$ within it \cite{pukhov}. As a result, laser energy mostly moves forward towards the front of the caviton, where it is rapidly redshifted and thus slips quickly back within the wave frame. This can cause pulse steepening at the \emph{front} of the pulse, and reduced steepening at the rear. 

In any case, the outcome of these complex dynamics is generally compression of the laser pulse and possible increase in $a^{2}$. Along with the exact pulse shape, this will determine $F_{p}$ and thus wakefield growth. Pulse compression of laser pulses in a laser wakefield has been reported previously \cite{FaurePRL2005}. Compression of a 30 fs pulse to $\sim$10 fs, was measured with a second-order autocorrelator. However, this method does not give phase information and thus does not allow a complete diagnosis of the pulse shape.

\indent In this letter we report on the complete temporal
(amplitude and phase) characterisation  of short ($\tau\simeq
45$\,fs), relativistically intense  ($I\approx 2\times
10^{19}$\,Wcm$^{-2}$) laser pulses after interaction with dilute
plasmas such that the initial pulse was shorter than the excited plasma wave wavelength ($c\tau < \lambda_{p}$).
Spectral broadening, photon acceleration and asymmetric
pulse shortening have been observed. The dependence of these
nonlinear effects on plasma density and interaction length have
been investigated, detailing the first parametric investigation of temporal
pulse evolution in a laser wakefield.

\indent The experiment was conducted with the Astra-Gemini laser at the
Rutherford Appleton Laboratory delivering laser pulses at central wavelength $\lambda_L=800\,$nm with energies up to
$12\,$J in a FWHM duration of $\tau_{0} = 45\,$fs.
The $200\,$TW laser pulses were focused by a $f/20$ off-axis
parabolic mirror to a spot size of ($w_0=22.0 \pm 0.6)\,\mu$m FWHM, with corresponding confocal parameter of $z_R = 1\,$mm. For these parameters, $a_{0} \approx 3.0$ in vacuum. 
The interaction with a helium gas jet with electron densities $n_{e}=1.2-6.6\times 10^{18}\,$cm$^{-3}$ and three different lengths
$l=4$, $6$ and $8.5\,$mm were investigated. $n_{e}$ and $l$ were determined by interferometry with a transverse optical probe \cite{KneipPRL2009}. Under these conditions the intensity modification was not sufficient to produce electron beams via self-injection.

The transmitted laser pulses were
collimated by a large aperture ($f/10$) spherical mirror and imaged onto a
12-bit CCD camera to confirm the quality of self-guiding.
The full beam was also focused onto the
slit of an optical spectrometer sensitive over 300-1000 nm, that was absolutely calibrated using a 
white light source. An absolutely calibrated photodiode measured
the transmitted laser energy. The temporal dependence of intensity
and phase were measured using second-order frequency resolved
optical gating (FROG) based on second harmonic generation (SHG) by guiding a small part of the
transmitted laser (close to the centre of the beam) into a SHG-FROG device
(GRENOUILLE by Swamp optics \cite{grenouille}) with $12\,$fs temporal and $4\,$nm spectral resolution. The amount of glass in the beam, such as vacuum windows, reflective neutral
density filters and the long focal length lens for recollimation,
was kept to a minimum and did not exceed $4.5\,$mm in total.

In SHG-FROG, a second-order autocorrelation is spectrally dispersed in the direction vertical to the time axis. The resulting 2D map - the FROG trace - thus contains both amplitude and phase information of the laser pulse. This allows the pulse shape to be uniquely determined, which is impossible from the autocorrelation alone. The pulse shape is established using a multiparametric fitting algorithm, referred to as retrieval (FROG3 \cite{FROG3}). In the retrieval, the temporal axis is discretised and, starting from an initial guess, the amplitude and phase of the laser at each point are varied until the measured FROG trace is best fitted by the synthetic one. The error is defined as the square root of the sum of the squared differences in each pixel of the measured and synthetic FROG. 

Experimentally obtained SHG-FROG traces can be asymmetric with respect to the time axis due to unwanted spatial chirp and pulse front tilt. Although we did not observe appreciable
asymmetries, the FROG traces were symmetrised prior to the
retrieval in order to improve the contrast and access a larger data set.
The analysis was restricted to retrievals with an
error smaller than $2\%$ 
and 
for which the total spectrum
measured in the spectrometer was in good agreement with the
retrieved spectrum. This gives confidence that the FROG results
are representative of the complete pulse and that no limitation due to
spectral clipping occurred. Finally, fields were corrected for
propagation through glass, vacuum window and filters in the beam
path. 
The pulse was attenuated by reflection off glass wedges before passing through any material so that only linear dispersion in the optics need be accounted for.

The complex laser fields are most usefully displayed in the form of a
Wigner-distribution $W(t,\,\omega)$ \cite{wigner} which represents the distribution of photons in
$(\omega,t)$ phase space. The marginals of the Wigner distribution, $\int
W(t,\,\omega)\textrm \, \mbox{d}\omega$ and $\int
W(t,\,\omega) \, \textrm{dt}$, give the temporal intensity and
spectrum of the pulse and its first normalised moment gives the
instantaneous frequency. 
Therefore, in a single 2D map it is
possible to visualise the most relevant information about the pulse.

\begin{figure}[b]
    \includegraphics[clip,width=8.5cm]{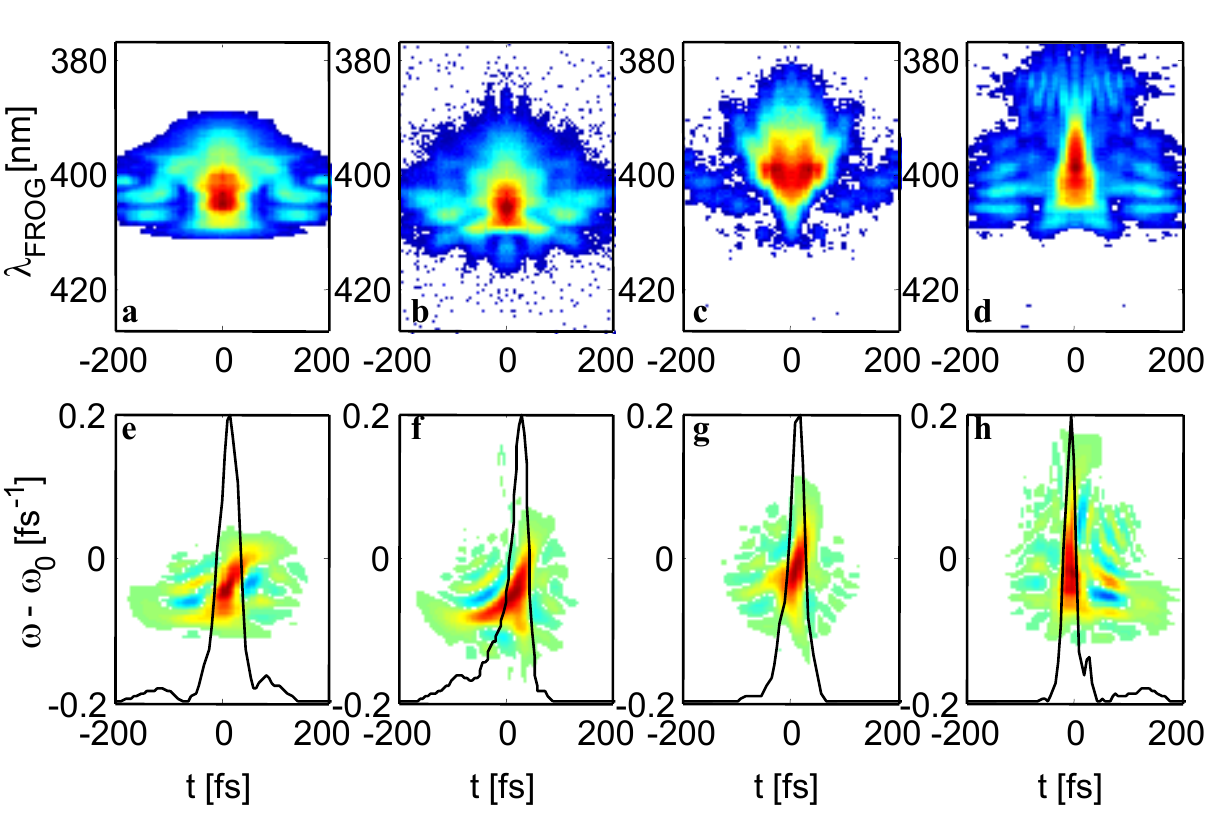}
    \caption{(color online) Length scan at low electron densities $n_{e} \in (1.2..2.3)\times 10^{18}$\,cm$^{-3}$. FROG-traces (first row) and
    Wigner-distributions of retrieved and corrected fields
    (second row) for $l = 0, 4, 6$ and $8.5\,$mm from left to right. Overlaid black lines correspond to temporal intensity distribution. $t<0$ represents front of the pulse.
    \label{fig1}}
    \end{figure}

\indent The remaining uncertainty of the direction of time due to
the symmetric nature of the SHG-FROG was removed using
physical considerations. First, a FROG trace of the pulse was obtained without gas, but with an \emph{additional} $4\,$mm of glass in its path (Fig.~\ref{fig1}a). The artificially introduced positive
chirp (increasing frequency with time) increases the total chirp measured (including the effect of the fixed glass in the system). 
Fig.~\ref{fig1}e shows the corresponding Wigner distribution after correcting for the glass in the beam path. We obtain an essentially bandwidth limited reference pulse with its temporal profile shown in black. The wings in the temporal profile are a common artefact in high power ultrashort laser systems and has little influence on the physical processes to be described.

Any similar positive dispersion due to the interaction would also increase the total chirp. Negative dispersion however would counter the positive chirp inherent in the system. So despite leading to the same pulse duration after the interaction, this would result in a quite different total chirp in the diagnostic. 
By varying the experiment parameters in small steps,  so that the transition from the
previous conditions were smooth, the change in the chirp and therefore the direction of
time could be determined for each retrieved pulse.

The pulse was measured after propagation through a plasma with relatively constant density $n_{e}\in1.2..2.3\times 10^{18}\,$cm$^{-3}$, low enough that the laser ($c\tau\sim13.5\, \mu$m) did not extend beyond the first plasma wave period ($\lambda_p \in 30..22\,\mu$m). For the shortest propagation distance $l=4\,$mm (Fig.~\ref{fig1}b,f) we observe a positive chirp as expected. Noticeable though is the faster steepening at the back of the pulse (at positive times). The front (negative times) stays comparable to the initial pulse and the total pulse-length is reduced. 
For a longer distance of $l = 6\,$mm (c,g), the front is depleted and for the longest distance $l=8\,$mm (d,h) the pulse becomes near-transform limited with a markedly reduced pulse duration ($\tau = 27 \pm 3\,$fs).

\begin{figure}[t]
    \includegraphics[width=8.6cm]{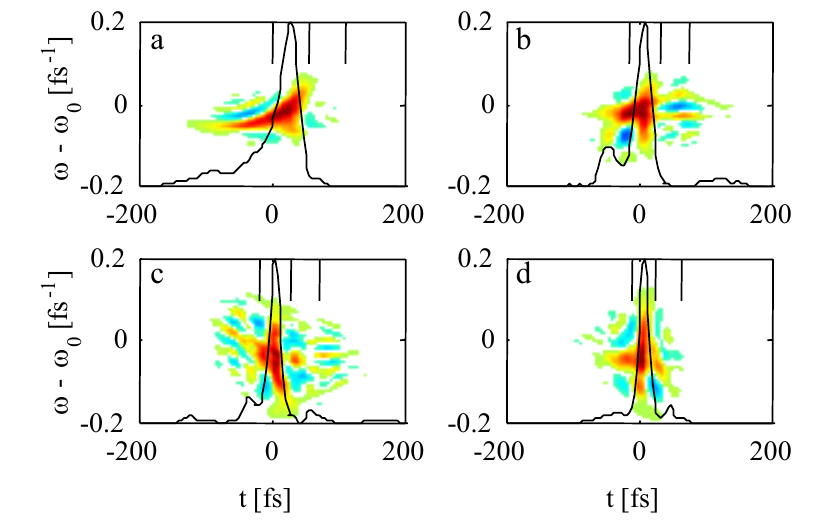}
    \caption{(color online) Pulse profiles for $n_{e} = $ (a) $2.3$, (b)
    $3.8$, (c) $4.7$, and (d) $6.6\times 10^{18}$cm$^{-3}$ for $l=4\,$mm. The black vertical
    lines mark the plasma period ($2\pi/\omega_p$), i.e.~expected positions of
    electron density maxima for the respective densities.
    \label{fig2}}
    \end{figure}

\indent Similar compression is observed for increasing $n_{e}$ for fixed propagation distance. Fig.~\ref{fig2} shows the Wigner-distributions and pulse profiles for the shortest interaction length ($l=4\,$mm), but this time for electron density increasing from $2.3$
to $6.6 \times 10^{18}$\,cm$^{-3}$ ($\lambda_p \in 22..13 \, \mu$m). The lowest density $n_{e}=2.3 \times 10^{18}\,$cm$^{-3}$ (Fig.~\ref{fig2}a) corresponds to the situation described in Fig.~\ref{fig1}b,d. With increasing $n_{e}$, the pulses show increased bandwidth and shorter pulse duration.  In contrast to the length scan at low density, the red-shift at the leading edge of the pulse is now more prominent. At $n_{e}=6.6 \times 10^{18}$\,cm$^{-3}$, the pulse is again almost transform limited with a broadened spectrum that supports the pulse duration $\tau = 18 \pm 3\,$fs. The pulses also exhibit enhanced structure with increasing density, with additional pre- and post-pulses observed. As a guide, the duration of one plasma period is indicated by the vertical lines
at the top of each panel in fig.~\ref{fig2}. The appearance of multiple
pulses and pulselets indicates the modulation of wings of the initial pulse by the plasma wave.

\begin{figure}
    \includegraphics[width=8.0cm]{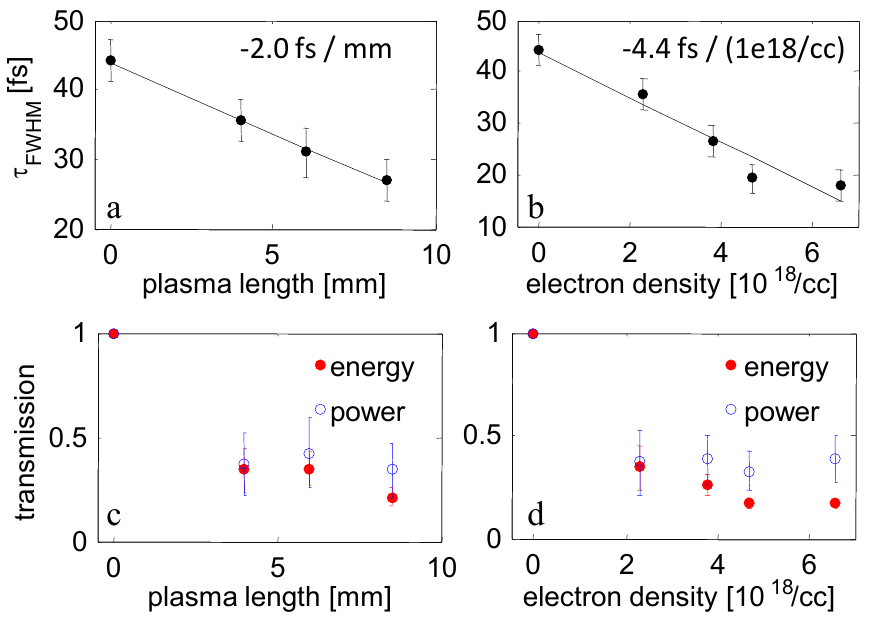}
    \caption{(color online) Dependence of pulse duration $\tau$ on (a) propagation length $l$ for
    $n_e=(1.2..2.3)\times 10^{18}$cm$^{-3}$ and (b) electron density
    $n_e$ for $l=4\,$mm. Best fit lines give compression
    rates of $2.0\,$fs/mm and $4.4\,$fs/($10^{18}$cm$^{-3}$) respectively.
    (c), (d) shows transmission in terms of laser
    energy (full circles) and peak power (open).
    \label{fig3}}
    \end{figure}

\indent Fig.~\ref{fig3}a, b quantify the pulse shortening. For constant density $n_{e}\sim1.8\times10^{18}$cm$^{-3}$ the FWHM duration reduces linearly from $\tau = 44\,$fs to $27\,$fs when the plasma length is stepwise increased to $l=8.5\,$mm. The rate of compression is $\approx2.0\pm 0.7\,$fs/mm. Similarly,
pulses are shortened from $44\,$fs to $18\,$fs for a constant $l=4\,$mm and $n_e$ increased from $2.3$ to $6.6 \times 10^{18}$cm$^{-3}$ giving a constant compression rate $\approx4.4\pm^{1.8}_{1.2}\,$fs/($10^{18}$cm$^{-3}$).

\indent Fig.~\ref{fig3}c,d show the variation in energy transmission, $E_{L}$ obtained from two complementary measurements; the integrated spectrum and the absolutely calibrated diode. ${E}_{L}$ decreases, 
with increasing propagation length and electron density. 
Denoting  the normalised temporal pulse profile as $f(l,n_e,t)$ ($l=n_e=0$ being the reference pulse), the transmitted energy, $E_L(l,n_e) = \int P_p(l,n_e) f\left(l,n_e,t\right) dt$, can then be used to calculate the relative peak power

    \begin{equation}\label{eq1}
    \frac{P_p(l,n_e)}{P_p(0,0)} = T\left(l,n_e\right) \cdot \frac{\int f\left(l,n_e,t\right) dt}{\int f\left(0,0,t\right) dt}
    \end{equation}
where $T(l,n_e)=E_L(l,n_e)/E_L(0,0)$ is the energy transmission. Fig.~\ref{fig3}c,d show a substantial  initial drop in $P_{p}$  
to $\approx30\,\%$ of its original value for $l=4$ mm. This can be attributed to the initial self-focusing, which traps only a fraction of the original laser energy in the matched guided pulse, especially for non-ideal beams \cite{thomas}. However for further propagation, the peak power changes only slowly over the range of parameters investigated. Evidently, under these conditions, pulse compression compensates for the energy loss of the pulse in continuously driving the wakefield. 

The pulse compression can be understood by considering two points in the pulse separated by $c \tau$ having different group velocities $v_{g1}$ and $v_{g2}$. Their separation therefore changes after propagating in $z$ direction over a distance $\Delta l$ according to $c\Delta \tau / \Delta l = (v_{g2} - v_{g1})/c = (\eta_{2} - \eta_{1}) \approx c\tau_{0} \frac{\partial \eta}{\partial z}$, where $\tau_{0}$ is the initial pulse duration. As we will show, as a reasonable first approximation, $\frac{\partial \eta}{\partial z}$ can be considered to be constant over the part of plasma period in which the majority of the laser energy sits for a pulse of duration $c\tau \approx \lambda_{p}$. Hence $\frac{\partial \eta}{\partial z} \approx (\eta_{min} - \eta_{max}) / c\tau$. For underdense plasma  $n_0 \ll n_c$, we have $\eta \approx 1-\frac{1}{2}(n_{e}/\gamma n_{c})$, where $n_{c}=m\varepsilon_0\omega^2/e^2$ is the critical density. Hence for sufficiently intense pulses; $a > 1$; $\eta_{max}\approx 1$; $\eta_{min} \approx 1- (n_{e0}/2n_{c})$. This gives a variation of pulse duration on initial density $n_{e0}$ and $l$, $\displaystyle \tau = \tau_0 - \frac{n_{e0} l}{2 c n_c}$.
For the two scans performed, we find compression rates $\Delta \tau/\Delta l = -n_{e0}/2cn_c = -1.7\,$fs/mm for $n_{e0} = 1.8 \times 10^{18}\,$cm$^{-3}$, and $\Delta \tau/\Delta n_0 = -l/2cn_c = - 3.8\,$fs/$10^{18}$cm$^{-3}$ with $l = 4\,$mm, in good agreement with the measured rates shown in Fig.~\ref{fig3}. 

\begin{figure}
    \includegraphics[width=8.0cm]{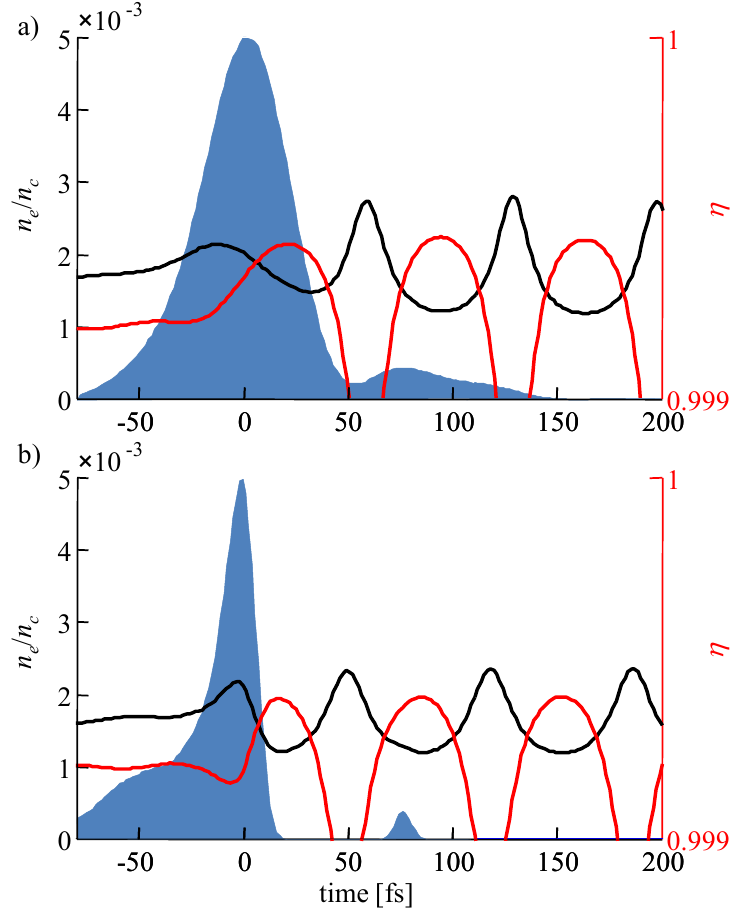}
    \caption{(color online) Numerical solution of 1D wave equation in quasistatic approximation. $a^{2}$ (blue filled), $n_{e}$ (black solid), and $\eta$ (red) resulting from propagation of a pulse with $a_0 \simeq 1.5$ at $n_{e0}=2.5\times 10^{18}$cm$^{-3}$; (a) initially, (b) at  $l=8$ mm. The reduced $a_0$ was chosen to account for the initial energy loss in the guided mode as in Fig.~\ref{fig3}.
    \label{fig5}}
    \end{figure}

The compression of a laser pulse in a laser wakefield has been studied through simulation in 1 and 2D for similar parameters by Gordon \emph{et al} \cite{gordon}. These simulations display many of the same traits as our observations, in particular, reproducing a positive chirp for short propagation distance at low density. 
 However to gain better insight into the pulse evolution, we consider the
1-dimensional quasistatic wave-equation for a pulse driving a wakefield \cite{sprangle}. Writing the normalised vector potential
$a_L$ in terms of a slowly varying envelope and a carrier frequency
term $k_0=\omega_0/c$, $a_L(\xi,\tau) = \frac{1}{2} a(\xi, \tau)\exp(ik_0
\xi) + c.c.$, the wave equation for the complex laser envelope reads
    \begin{equation}\label{eq2}
    2\frac{\partial^2a}{\partial\xi\partial\tau}+2ik_0\frac{\partial a}{\partial\tau} - \frac{\partial^2a}{\partial\tau^2}=\frac{a}{1+\Phi} ,
    \end{equation}
where $\xi=z-ct$ and $\tau=t$ are the spatial and temporal coordinates in
the comoving frame. $\Phi$ is the electrostatic potential in
the plasma, which is self-consistently obtained by solving Poisson's equation
\cite{sprangle}
    \begin{equation}\label{eq3}
    \frac{\partial^2 \Phi}{\partial\xi^2} = \frac{1}{2}\left[
    \frac{\gamma^2}{(1+\Phi)^2}-1\right] \ ,
    \end{equation}
where $\gamma^2=1+|a|^2/2$ as before.
All quantities in eqs.~(\ref{eq2}) and (\ref{eq3}) are normalised
to the characteristic plasma parameters, i.e.~space to $c/\omega_p$ and time to $1/\omega_p$.
The intensity distribution $a^2$ and the refractive index $\eta$ obtained from the numerical integration of eqs.~(\ref{eq2}) and (\ref{eq3}) is shown in Fig.~\ref{fig5}.

The model shows that, at low densities, pulse compression starts  from the back of the pulse, as observed experimentally. At $l=8$ mm, the pulse has compressed to $\tau\approx 21\,$fs, in good agreement with our measured compression. Laser energy which sits in following buckets is trapped to form trailing pulses. By contrast the front of the pulse evolves only slowly. Since for relativistic pulses it is the front edge of the pulse which determines the wakefield amplitude, this also evolves slowly. Hence in this regime (low $n_{e}$, high $a$), pulse compression can extend the growth of the plasma wave to distances well beyond a naive depletion 
length obtained by considering only energy transfer to the wake. Furthermore here, depletion certainly cannot be modelled by pulse front erosion \cite{lu,bulanov}.

At these densities, there is little evidence for the explosive increase in $a^{2}$ predicted for propagation longer than the non-linear modification time \cite{gordon}, which can be stated simply as, $l> c\tau_{nl} \approx (c\tau_{0}/|a|^{2})(n_{c}/n_{e})$ \cite{bulanov}. For $n_{c}/n_{e}\sim 1000$, and our initial pulse length, $c\tau_{nl}> 10$ mm, and for $l=4$ mm, this time is not reached for $n_{e} < 6\times10^{18}\, \rm{cm}^{-3}$, \emph{provided} in both cases that $|a|^{2}$ is not too much greater than 1. This supports the measurement shown in figure 3, that for our conditions, only a fraction of the laser energy is captured in the wakefield driving filament beyond $z_{R}$. This also explains the good agreement with the 1D calculations and also explains why there is not sufficient plasma wave growth to observe self-injected electron beams below this density. Pulse evolution leading to $a^{2}$ amplification and the resultant wavebreaking of the plasma wave has been inferred from the properties of generated electron beam at the high end of this density range but over longer interaction lengths  \cite{KneipPRL2009}.  Though challenging, future studies into this non-linear regime of pulse compression, which may produce extremely short laser pulses of higher power, may prove to be particularly rewarding.

\begin{acknowledgments}
\indent The authors would like to thank the staff of the Central Laser Facility for their assistance during the experiments and A.~E.~Dangor for useful discussions. JS acknowledges financial support from DAAD.
\end{acknowledgments}

\noindent $^{\dag}$ Present address: University of California San Diego, USA \\
$^{\ddag}$ Present address: Technical Institute of Crete, Greece\\

\end{document}